\documentclass[9pt,twocolumn,twoside]{opticajnl}
\journal{opticajournal} 

\setboolean{shortarticle}{true}

\title{Robust Anderson transition in non-Hermitian photonic quasicrystals}

\author[1,2,*]{Stefano Longhi}

\affil[1]{Dipartimento di Fisica, Politecnico di Milano, Piazza L. da Vinci 32, I-20133 Milano, Italy}
\affil[2]{IFISC (UIB-CSIC), Instituto de Fisica Interdisciplinar y Sistemas Complejos - Palma de Mallorca, Spain}

\affil[*]{stefano.longhi@polimi.it}

\begin{abstract}
Anderson localization, i.e. the suppression of diffusion in lattices with random or incommensurate disorder, is a fragile interference phenomenon which is spoiled out in the presence of dephasing effects or fluctuating disorder. As a consequence, Anderson localization-delocalization phase transitions observed in Hermitian systems, such as in one-dimensional quasicrystals when the amplitude of the incommensurate potential is increased above a threshold, are washed out when dephasing effects are included. Here we consider localization-delocalization spectral phase transitions occurring in non-Hermitian quasicrystals with local incommensurate gain and loss, and show that, contrary to the Hermitian case, the non-Hermitian phase transition is robust against dephasing effects. The results are illustrated by considering synthetic quasicrystals in photonic mesh lattices.  
\end{abstract}

\setboolean{displaycopyright}{false} 

\begin{document}

\maketitle


Anderson localization \cite{r1}, i.e. the absence of diffusion in systems with static disorder, is ubiquitous in the physics of disordered systems \cite{r2}. Photonics has provided  since more than three decades a fascinating laboratory platform for testing the localization properties of disordered systems, disclosing their rich physics \cite{r3,r4,r5,r6,r7,r8,r9,r10,r11,r12,r13} with potential applications to the design of the next generation of materials \cite{r14}.  In low-dimensional systems, the Anderson localization-delocalization transition is observed in quasicrystals, i.e. in lattices displaying aperiodic order \cite{r16}. The most notably example is provided by the Aubry-Andr\'e  model \cite{r17}. This model describes a one-dimensional incommensurate potential on a lattice, in which all wave functions in the spectrum suddenly change from extended to localized when the strength of the quasiperiodic potential exceeds a threshold \cite{r16,r17}. Such a phase transition has been experimentally observed in several physical systems, including photonic waveguide arrays and Bose-Einstein condensates in optical lattices \cite{r18,r19,r20}. Recently, a non-Hermitian (NH) extension of the Aubry-Andr\'e model, where the incommensurate potential is complex and accounts for aperiodic local gain and loss in the system, has been introduced \cite{r21,r22,r23,r24}, and the topological origin of the metal-insulator phase transition has been disclosed \cite{r21}, with experimental demonstrations reported in recent works \cite{r25,r26}.\\ 
Anderson localization is rather generally a fragile phenomenon against dephasing, decoherence or frequent measurements \cite{r6,r27,r28,r29,r30,r31,r32}:  dephasing or fluctuating potentials destroy interference effects, Anderson localization
is  spoiled out and replaced by diffusive transport \cite{r6}. This implies that metal-insulator phase transitions, such as those observed in the Hermitian Aubry-Andr\'e  models, are washed out by dephasing effects.  A natural and largely open question is whether Anderson localization is fragile or robust against dephasing when the underlying Hamiltonian is non-Hermitian.\par
In this Letter we address such a main question and show that the localization-delocalization phase transition in a specific quasicrystal model, described by the NH extension of the Aubry-Andr\'e model with dissipative disorder \cite{r21}, is robust against dephasing effects. The results are illustrated by considering synthetic quasicrystals realized in photonic mesh lattices \cite{r6,r26,r27,r37,r38,r39,r40}. It is also pointed out that non-Hermiticity itself, neither the topological nature of Anderson transitions in non-Hermitian models, are not sufficient conditions to ensure robustness against dephasing.\\
 To introduce the problem in a rather general framework, let us consider the coherent evolution of excitations in a tight-binding lattice, comprising $N$ sites and described by the matrix Hamiltonian $H=H_{n,m}$ ($n,m=1,2,...,N$), and let us assume that at every time interval $\Delta t$ the phase of the wave function $\psi_n$ at any lattice site $n$ is randomized. Such a randomized phase process basically emulates dephasing effects in the dynamics and has been used in photonic quantum walks to experimentally demonstrate transition from Anderson localization to diffusive transport for fluctuating disorder \cite{r6}. 
 The time evolution of the wave function amplitudes $\psi_n(t)$ reads
 \begin{equation}
 i \frac{d \psi_n}{dt}= \sum_m H_{n,m} \psi_m+ \psi_n \sum_{\rho=1,2,3,...} \varphi_n^{(\rho)} \delta(t-\rho \Delta t)
 \end{equation}
where the last term on the right hand sides of Eq.(1) describes the dephasing process. Here $\varphi_n^{(\rho)}$ are assumed to be uncorrelated stochastic phases, both in site index $n$ and time step $\rho$, with a given probability density function. Fully coherent dynamics is obtained by letting $\varphi_n^{(\rho)}=0$, whereas fully incoherent (classical) dynamics is obtained by assuming a uniform distribution in the range $(-\pi, \pi)$ for the probability density function. 
Under fully coherent dynamics, the wave function amplitudes evolve according to $ \psi_n(t)=\sum_m U_{n,m}(t) \psi_m(0)$, where $U(t)=\exp(-i Ht)$ is the coherent propagator. On the other hand, for fully incoherent dynamics indicating by $P_n(t_\rho)={\overline {|\psi_n(t_{\rho})|^2}}$ the occupation probabilities (populations) at various sites of the lattices, where $t_{\rho}=\rho \Delta t$ and the overbar denotes statistical average, the time evolution is described by the classical map
\begin{equation}
P_n(t_{\rho+1})=\sum_m \mathcal{U}_{n,m}P_m(t_{\rho})
\end{equation}
where $\mathcal{U}_{n,m}=|U_{n,m} (\Delta t)|^2$ in the incoherent propagator. Equation (2) can be readily obtained by solving Eq.(1) in each time interval $\Delta t$ and then taking the statistical average of $| \psi_n|^2$. It is instructive to unveil the fate of Anderson localization under dephasing. Let us assume that $H$ displays spectral Anderson localization. This means that the eigenstates of $H$ are exponentially localized, which corresponds generally (but not universally) to suppression of wave spreading  in the lattice (dynamical localization). Under fully incoherent dynamics, provided that the Hamiltonian $H$ is Hermitian it can be demonstrated  that, as expected, Anderson localization is spoiled out and excitation spreads in the lattice, finally reaching a stationary state with equal distributions of populations in the lattice, i.e. $P_n(t_{\rho} \rightarrow \infty)=1/N$. The proof is given in Sec.1 of the Supplemental document.
For example, in the Aubry-Andr\'e model describing a one-dimensional quasicrystal, the Hamiltonian $H$ reads explicitly
\begin{equation}
H_{n,m}=J(\delta_{n,m+1}+\delta_{n,m-1})+2V_0 \delta_{n,m} \sin ( 2 \pi \alpha n+ih)
\end{equation}
where $J$ is the hopping amplitude between adjacent sites in the lattice, $2 V_0$ is the amplitude of the incommensurate sinusoidal potential, $\alpha$ is irrational Diophantine, and $ih$ a complex phase term \cite{r21}. The Hermitian case is obtained by letting $h=0$. In this case under coherent dynamics the model displays a metal-insulator phase transition when $V_0$ is increased above the critical value $V_0=J$. Clearly, such a phase transition is not observed anymore in the incoherent regime (see Fig.S1 in the Supplemental Material). However, in the NH case $h \neq 0$ a different metal-insulator phase transition can be observed under coherent dynamics \cite{r21}, where the control parameter is provided by the complex phase $h$. Assuming $V_0<J$, for $h=0$ the system is in the delocalized phase. As $h$ is increased above the critical value 
\begin{equation}
h_c= \log (J/V_0), 
\end{equation}
the system undergoes a delocalization-localization phase transition, with all eigenstates being exponentially localized for $h>h_c$ \cite{r21}. Remarkably, and this is the central result of this work, such a NH phase transition {\em is not} washed out by dephasing effects, contrary to what happens for Hermitian phase transitions. This point is discussed further in Sec.2 of the Supplemental document. Why is localization robust in this case? Basically in the Hermitian quasicrystal, where the incommensurate on-site potential $V_n$ is real, dephasing effects spoil the delicate phase interference of waves scattered off by the potential disorder, which is at the heart of Anderson localization. However, when the on-site potential is a complex function, i.e. when we have lattice disorder in local gain or loss, phase randomization does not affect the local damping or amplification of the wave function, and such a dissipative disorder can provide a route toward localization robust against dephasing effects. Specifically,
 assuming a small potential amplitude $V_0/J \sim \epsilon$, with $\epsilon$ a small parameter, and a short time interval $\Delta t  \sim \epsilon /J$, under incoherent dynamics a delocalization-localization phase transition  is observed as $h$ is increased above the critical value
\begin{equation}
h^{'}_{c}= {\rm asinh} \left( \frac{\Delta t J^2}{2 V_0} \right). 
\end{equation}
 \begin{figure}[ht]
 \centering
    \includegraphics[width=0.48\textwidth]{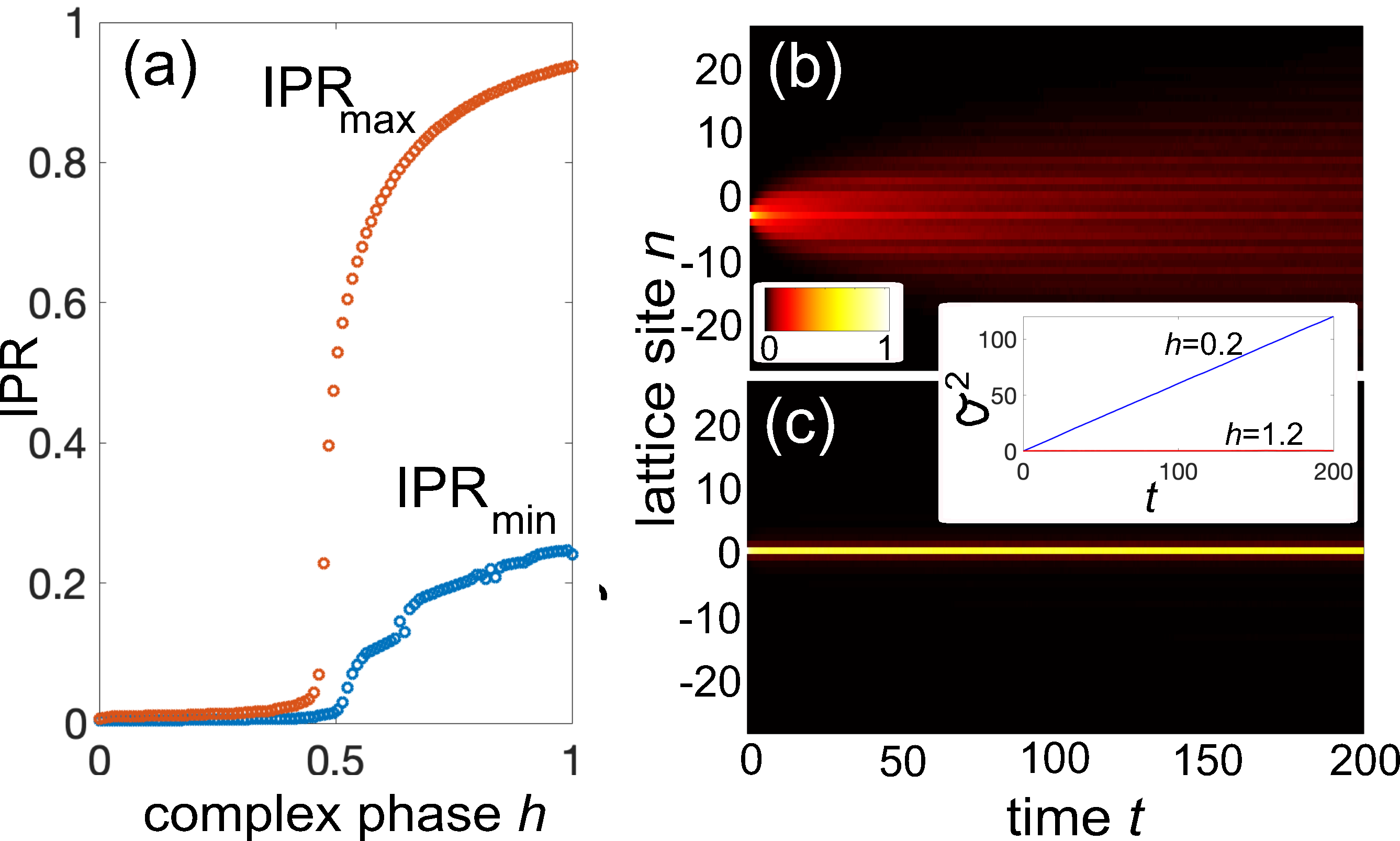}
   \caption{ \small (a) Behavior of the maximum and minimum inverse participation ratio (IPR) of eigenstates of the incoherent propagator $\mathcal{U}$ versus the complex phase $h$ in the NH Aubry-Andr\'e model for parameter values $J=1$, $V_0=0.3$, $\Delta t=0.3$ and $\alpha=(\sqrt{5}-1)/2$. (b) Temporal evolution of the normalized occupation probability $P_n(t)$ versus time $t$ in the delocalized phase ($h=0.2$). The excitation spreads diffusively in the lattice. (c) Same as (b) but for $h=1.2$, corresponding to localization. The inset in panels (b,c) depicts the corresponding temporal evolution of the second moment $\sigma^2(t)$ in the two cases.}
\end{figure}
An example of the incoherent NH phase transition is shown in Fig.1. The localization properties of $\mathcal{U}$ are characterized by the inverse participation ratio (IPR) of any eigenvector $\Theta_n$ of $\mathcal{U}$, which for a normalized eigenstate is defined by ${\rm IPR}= \sum_n | \Theta_n|^4$. 
For a tightly-confined eigenstate, the IPR is independent of $N$ and takes a finite value close to 1, whereas for an extended state the IPR is small and vanishes as $ \sim 1/N$ as $N \rightarrow \infty$. The largest and smallest values of the IPR for any eigenvector of $\mathcal{U}$ are indicated by IPR$_{max}$ and IPR$_{min}$, respectively. Figure 1(a) shows the behavior IPR$_{max}$ and IPR$_{min}$ versus $h$ for parameter values $\alpha= (\sqrt{5}-1)/2$, $J=1$, $V_0=\Delta t=0.3$, clearly indicating a delocalization-localization phase transition near the critical point $h_c^{'} \simeq 0.48$ as predicted by Eq.(5). The eigenstates of $\mathcal{U}$ have been numerically computed for a lattice size $N=233$ under periodic boundary conditions, and using the rational approximant $\alpha= 144/233$ of the inverse of the golden ratio \cite{r21}. 
The distinct transport properties, in the delocalized ($h<h_c^{'}$) and localized ($h>h_c^{'}$) phases, are shown in Figs.1(b) and (c). The figure panels illustrate the temporal evolution of occupation probabilities $P_n(t)=\overline{ | \psi_n(t)|^2}$, normalized at each time step such that $\sum_n P_n(t)=1$, as obtained by numerical integration of the Schr\"odinger equation (1), after a statistical average over 1000 realizations of stochastic phase randomization. The lattice is initially excited in site $n=0$. The spreading dynamics is characterized by the second moment $\sigma^2(t)=\sum_n P_n(t) n^2 / \sum_n P_n(t)$, which is depicted in the inset of Figs.1(b,c). Note that in the delocalized phase [Fig.1(b)] the second moment increases linearly with time $t$, corresponding to diffusive spreading, whereas in the localized phase [Fig.1(c)] wave spreading is halted.\\
It should be mentioned that the robustness of Anderson transition against dephasing strictly requires some dissipative disorder, and thus it is not observed in other NH models. For example, in a quasicrystal with real on-site potential and asymmetric hopping rates, like in the Hatano-Nelson model, the localization-delocalization transition of topological nature is not robust against dephasing (see Sec.3 of the Supplemental material).\\
 To illustrate the predicted phenomenon in an experimentally accessible platform, let us consider a photonic implementation of a quasicrystal based on light pulse dynamics in synthetic mesh lattices \cite{r6,r26,r27,r41}, where decoherence can be introduced and controlled by random dynamic phase changes \cite{r6}. The system  consists of two fiber loops of slightly different lengths that are connected by a fiber coupler with a coupling angle $\beta$. Phase and amplitude modulators are placed in one of the two loops, which provide a desired control of the phase and amplitude of the traveling pulses.  Light dynamics is described by the set of discrete-time equations \cite{r26,r38,r39,r41}
 \begin{eqnarray}
 u^{(m+1)}_n & = & \left(   \cos \beta u^{(m)}_{n+1}+i \sin \beta v^{(m)}_{n+1}  \right)  \exp (-2i\phi_{n}^{(m)}) \; \\
 v^{(m+1)}_n & = & \left(   \cos \beta v^{(m)}_{n-1}+i \sin \beta u^{(m)}_{n-1}  \right)
 \end{eqnarray}
 where $u_n^{(m)}$ and $v_n^{(m)}$ are the pulse amplitudes at discrete time step $m$ and lattice site $n$ in the two fiber loops, and $2 \phi_n^{(m)}$ comprises the phase and amplitude changes impressed by the modulators. The modulators are driven such that
 \begin{equation}
 \phi_{n}^{(m)}=2V_0 \sin (2 \pi \alpha n+ih)+\varphi_n^{(m)}
 \end{equation}
  \begin{figure}[ht]
 \centering
    \includegraphics[width=0.48\textwidth]{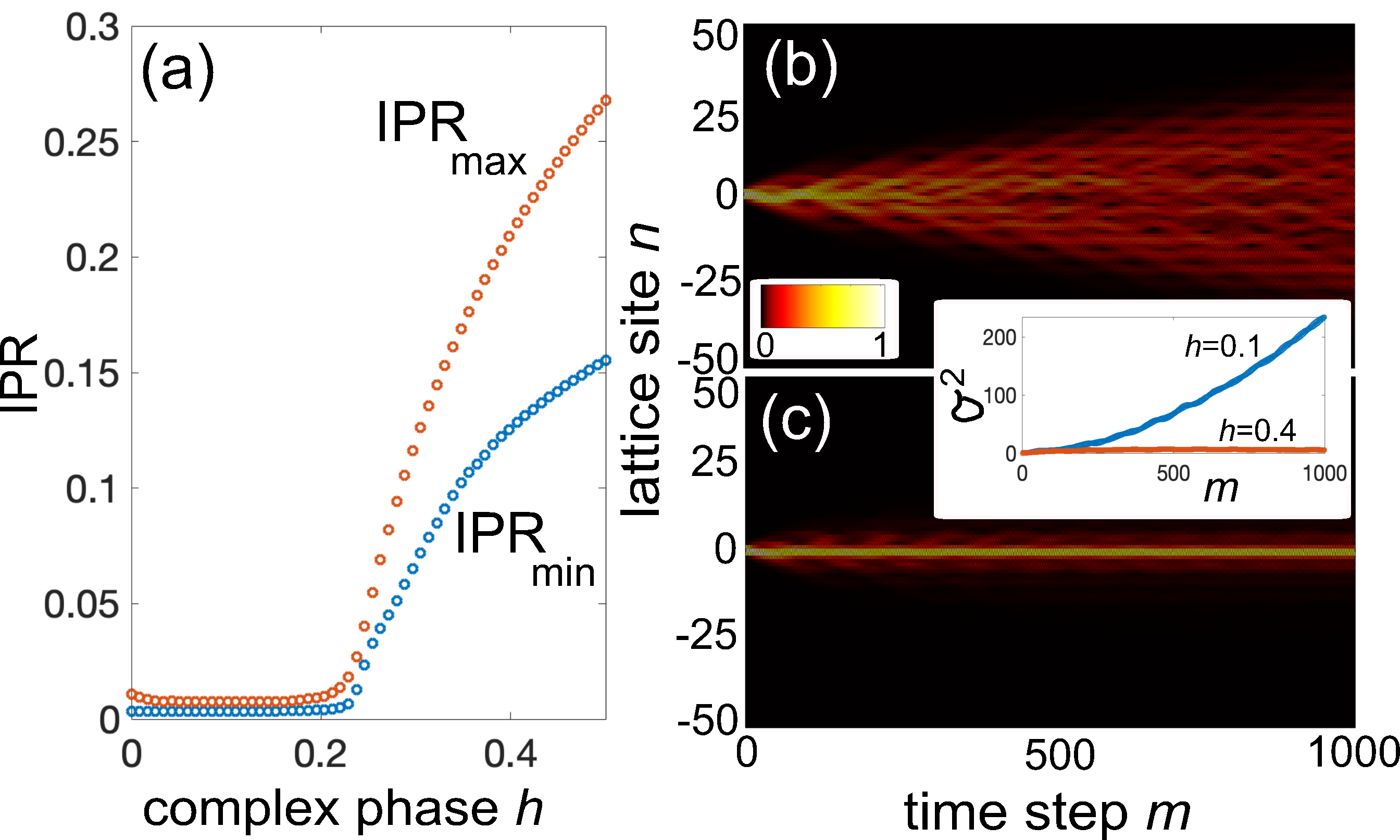}
   \caption{ \small Non-Hermitian localization-delocalization phase transition in a photonic quasicrystal under coherent dynamics. (a) Behavior of IPR$_{min}$  and IPR$_{max}$ of eigenstates of the coherent propagator versus the complex phase $h$  for parameter values $\beta =0.96 \times \pi/2$, $V_0=0.025$, and $\alpha=(\sqrt{5}-1)/2$. Note the localization-delocalization transition at the critical value $h_{c} \simeq 0.228$, according to the theoretical prediction [Eq.(9)].  (b) Temporal evolution of the normalized occupation probability $P_n^{(m)}$ versus time step $m$ under coherent dynamics in the delocalized phase ($h=0.1$). The excitation spreads ballistically in the lattice, corresponding to a quadratic increase of $\sigma^2$ with time step $m$. (c) Same as (b) but in the localized phase  ($h=0.4$). The inset in panels (b,c) depicts the temporal evolution of the second moment $\sigma^2(m)$ versus time step in the two cases.}
\end{figure}
  where the first term on the right hand side of Eq.(8) describes the static incommensurate complex on-site potential of the quasicrystal \cite{r41} whereas the additional stochastic phases $ \varphi_n^{(m)}$, when applied, introduce decoherence in the system \cite{r6}. For a coupling angle $\beta$ close to $\pi/2$ and for a small potential amplitude such that $V_0 \exp(h) \ll 1$, under coherent dynamics, i.e. for $\varphi_n^{(m)}=0$, the model described by Eqs.(6) and (7) reproduces the NH Aubry-Andr\'e model [Eq.(3)] with an hopping amplitude $J= \pm (1/2) \cos \beta$ and on-site potential $V_n= \phi_n=2 V_0 \sin (2 \pi \alpha n +ih)$ (see Sec.3 of the Supplemental document; see also \cite{r41}). Assuming $V_0<J$, a phase transition should be therefore observable as $h$ is varied to cross the critical point $h=h_c$, where 
  \begin{equation}
  h_c=\log \left( \frac{ \cos \beta} { 2 V_0} \right).
  \end{equation}
  The theoretical prediction is confirmed by full numerical simulations of Eqs.(6) and (7). Typical numerical results, clearly showing a delocalization-localization phase transition as the complex phase $h$ is increased above the critical value $h_c$, are shown in Fig.2. Figure 2(a) shows the behavior of maximum and minimum IPR of eigenstates of the one-step coherent propagator versus the complex phase $h$.
   A dynamical fingerprint of the NH phase transition is the dynamical delocalization of the wave packet for $h<h_c$ and the dynamical localization for $h>h_c$, as shown in Figs.2(b) and (c). The two panels illustrate the discrete-time evolution of light intensity distribution in the lattice, $P_n^{(m)}=|u_n^{(m)}|^2+|v_n^{(m)}|^2$, normalized at each time step ($P_n^{(m)} \rightarrow P_n^{(m)} / \sum_n P_n^{(m)}$), when a single pulse is injected into the system at lattice site $n=0$, namely for the initial condition $u_n^{(0)}=v_n^{(0)}=(1/ \sqrt{2}) \delta_{n,0}$. The spreading of the light wave packet in the lattice at successive time steps is measured by the second moment 
$  \sigma^2(m)= \sum_n n^2 P_n^{(m)} / \sum_n P_n^{(m)}$. The numerical results clearly demonstrate dynamical delocalization for $h<h_c$ with ballistic transport ($\sigma^2$ increases quadratically with time step $m$)  and dynamical localization for $h>h_c$.\\
Under incoherent dynamics, i.e. when $\varphi_n^{(m)}$ are uncorrelated stochastic phases with uniform distribution in the range $(-\pi, \pi)$, the incoherent light evolution is described by the following map for the light pulse intensities $X_n^{(m)}=\overline{|u_n^{(m)}|^2}$ and $Y_n^{(m)}=\overline{|v_n^{(m)}|^2}$ in the two fiber loops
\begin{eqnarray}
X_n^{(m+1)} & = & \left( \cos^2 \beta X_{n+1}^{(m)}+ \sin^2 \beta Y_{n+1}^{(m)} \right) \exp(g_n) \\
Y_n^{(m+1)} & = & \sin^2 \beta X_{n-1}^{(m)}+ \cos^2 \beta Y_{n-1}^{(m)} 
\end{eqnarray}
 where the overline denotes statistical average and where we have set
 \begin{equation}
 g_n=4 \;{\rm Im}(\phi_n^{(m)})=8 V_0 \sinh h \cos(2 \pi \alpha n). 
 \end{equation}
 The above equations describing incoherent light dynamics are readily obtained after taking the modulus square of both sides in Eqs.(6) and (7) and making the statistical average, using the property that $\overline{u_n^{(m)} v_n^{(m)*}}=0$ owing to phase randomization. In the Hermitian limit $h=0$, i.e. for $g_n=0$, Eqs.(10) and (11) describe a classical random walk, Anderson localization is washed out and transport in the lattice is diffusive \cite{r6}. In this case for a finite lattice of size $N$ with open boundaries, the dynamics is finally attracted toward the state with equal populations in the various sites, i.e. $X_n=Y_x=1/(2N)$.  As $h$ is increased, remarkably a delocalization-localization phase transition can be observed and transport in the lattice is suppressed. The critical value $h^{'}_c$ of the imaginary phase at which the phase transition occurs can be calculated analytically for a coupling angle $\beta$ close to $\pi/2$ and reads (see Sec.5 of the Supplemental document for technical details)
 \begin{equation}
 h^{'}_{c}= {\rm asinh} \left( \frac{\cos^2 \beta} {4 V_0} \right).
 \end{equation}
 \begin{figure}[ht]
 \centering
    \includegraphics[width=0.48\textwidth]{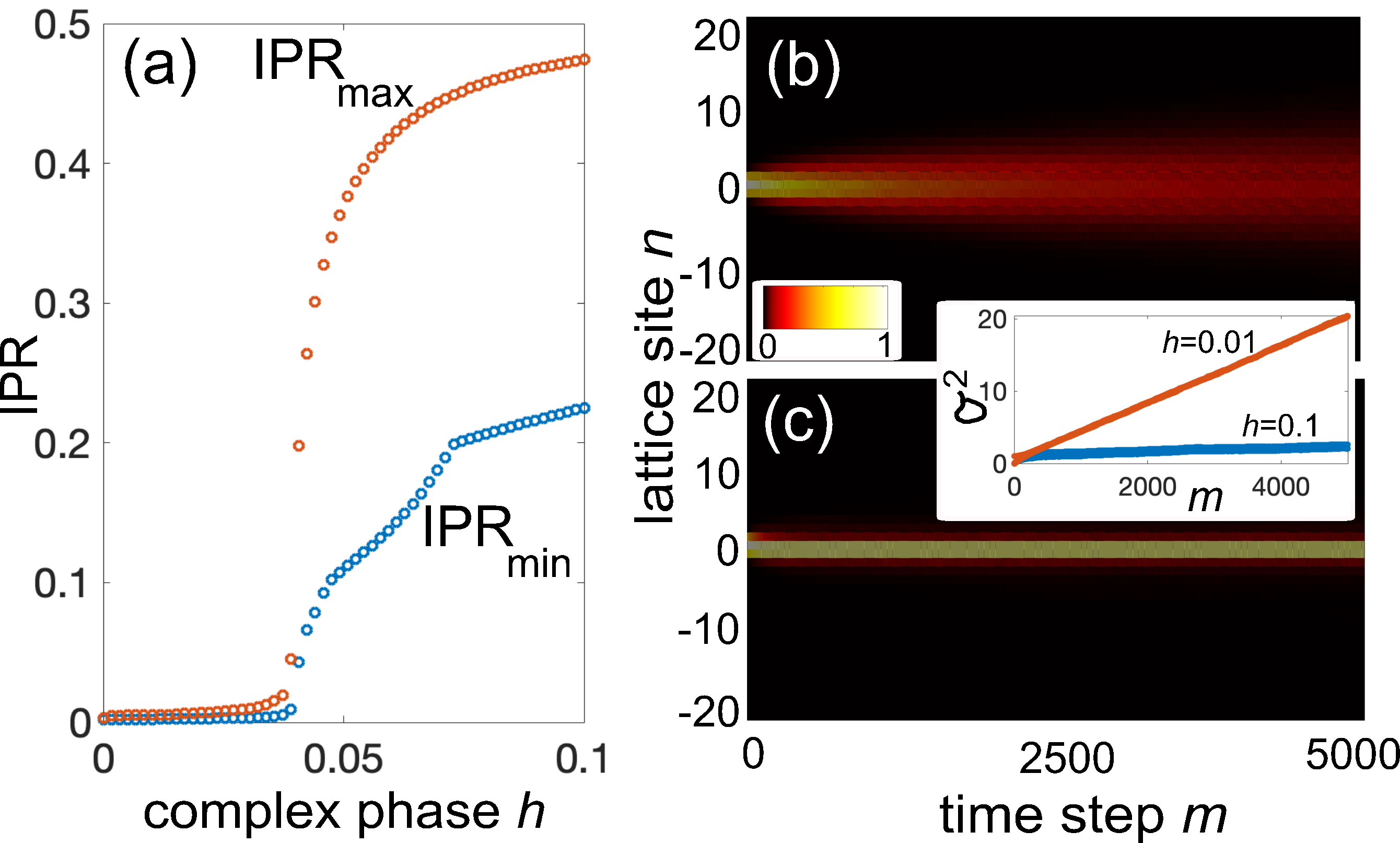}
   \caption{ \small Non-Hermitian localization-delocalization phase transition in a photonic quasicrystal under incoherent dynamics for the same parameter values as in Fig.2. (a) Behavior IPR$_{min}$ and IPR$_{max}$ of eigenstates of the incoherent propagator versus the complex phase $h$  for the same parameter values as in Fig.2. Note the localization-delocalization transition at the critical value $h_{c}^{'} \simeq 0.0394$, according to the theoretical prediction of Eq.(13).  (b) Temporal evolution of the normalized occupation probability $P_n^{(m)}$ versus time step $m$ under incoherent dynamics in the delocalized phase ($h=0.01$). The excitation spreads diffusively in the lattice, corresponding to a linear increase of the second moment $\sigma^2$ with time step $m$. (c) Same as (b) but in the localized phase  ($h=0.1$), The inset in panels (b,c) depicts the temporal evolution of the second moment $\sigma^2(m)$ versus time step $m$ in the two cases.}
\end{figure}
\noindent Typical numerical results, showing the delocalization-localization phase transition with dephasing effects as $h$ is increased above the critical value $h^{'
}_{c}$, are depicted in Fig.3. Note that for $h<h^{'}_{c}$ the system is in the delocalized phase and dynamical delocalization is observed, corresponding to diffusive wave spreading (contrary to ballistic spreading as in the coherent regime). On the other hand, spectral and dynamical localization are observed for $h>h^{'}_{c}$.

\noindent

 In conclusion, we  predicted that, while Anderson localization in Hermitian models is generally a fragile effect and is washed out in the presence of dephasing or fluctuating potentials, in non-Hermitian systems with dissipative disorder Anderson localization can survive against dephasing effects. We illustrated such a remarkable result by showing  robustness of localization-delocalization phase transitions in non-Hermitian quasicrystals with incommensurate local gain and loss, which should be observable in synthetic quasicrystals based on light pulse dynamics in coupled fiber loops.  Our results unravel  new physical insights onto Anderson localization and indicate that photonic quantum walks could provide an experimentally accessible platform for the observation of persistent Anderson localization in NH systems against dephasing effects.\\
\\
\noindent
{\bf Disclosures}. The author declares no conflicts of interest.\\
{\bf Data availability}. No data were generated or analyzed in the presented research.\\
{\bf Supplemental document}. See Supplement 1 for supporting content.

\newpage


 {\bf References with full titles}\\
 \\
 \noindent
1. P.W. Anderson, Absence of Diffusion in Certain Random Lattices, Phys. Rev. {\bf 109}, 1492 (1958).\\ 
2. A. Lagendijk, B. van Tiggelen, and D.S. Wiersma, Fifty years of Anderson localization, Physics Today {\bf 62}, (8) 24 (2009).\\
3. D. S. Wiersma, P. Bartolini, A. Lagendijk, and R. Righini,  Localization of light in a disordered medium, Nature {\bf 390}, 671 (1997).\\
4. T. Schwartz, G. Bartal, S. Fishman, and M. Segev, Transport and Anderson localization in disordered two-dimensional photonic lattices, Nature {\bf 446}, 52 (2007).\\
5. Y. Lahini, A. Avidan, F. Pozzi, M. Sorel, R. Morandotti, D.N. Christodoulides, and Y. Silberberg, Anderson Localization and Nonlinearity in One-Dimensional Disordered Photonic Lattices, Phys. Rev. Lett. {\bf 100}, 013906 (2008).\\
6. A. Schreiber, K. N. Cassemiro, V. Potocek, A. Gabris, I. Jex, and Ch. Silberhorn, Decoherence and Disorder in QuantumWalks: From Ballistic Spread to Localization, Phys. Rev. Lett. {\bf 106}, 180403 (2011).\\
7. U. Naether, Y.V. Kartashov, V.A. Vysloukh, S. Nolte, A. T\"unnermann, L. Torner, and A. Szameit,
Observation of the gradual transition from one-dimensional to two-dimensional Anderson localization, Opt. Lett. {\bf 37}, 593 (2012).\\
8. S. St\"utzer, Y. V. Kartashov, V. A. Vysloukh, A. T\"unnermann, S. Nolte, M. Lewenstein, L. Torner, and A. Szameit, Anderson cross-localization, Opt. Lett. {\bf 37}, 1715 (2012).\\
9. M. Segev, Y. Silberberg, and D.N. Christodoulides, Anderson localization of light,
Nature Photon. {\bf 7}, 197 (2013).\\
10.  A. Crespi, R. Osellame, R. Ramponi, V. Giovannetti, R. Fazio, L. Sansoni, F. De Nicola, F. Sciarrino, and P. Mataloni,
Anderson localization of entangled photons in an integrated quantum walk,  
Nature Photon. {\bf 7}, 322 (2013).\\
11. I.D. Vatnik, A. Tikan, G. Onishchukov, D.V. Churkin, and A.A. Sukhorukov, Anderson localization in synthetic photonic lattices, Sci. Rep. {\bf 7}, 4301 (2017).\\
12. M. Lee, J. Lee, S. Kim, S. Callard, C. Seassal, and H. Jeon, Anderson localizations and photonic band-tail states observed in compositionally disordered platform, Sci. Adv. {\bf 4}, e1602796 (2018).\\
13. A. Dikopoltsev, S. Weidemann, M. Kremer, A. Steinfurth, H.H. Sheinfux, A. Szameit, and M. Segev,
Observation of Anderson localization beyond the spectrum of the disorder, Sci. Adv. {\bf 8}, eabn7769 (2022).\\
14. S. Yu, C.-W. Qiu, Y. Chong, S. Torquato, and N. Park, Engineered disorder in photonics,
Nature Rev. Mat. {\bf 6}, 226 (2021).\\
15. J. B. Sokoloff, Unusual band structure, wave functions and electrical conductance in crystals with incommensurate periodic potentials, Phys. Rep. {\bf 126}, 189 (1984).\\
16. S. Aubry and G. Andr\'e, Analyticity breaking and Anderson localization in incommensurate lattices, Ann. Israel Phys. Soc {\bf 3}. 18 (1980).\\
17. Y. Lahini, R. Pugatch, F. Pozzi, M. Sorel, R. Morandotti, N. Davidson, Y. Silberberg, Observation of a localization transition in quasiperiodic photonic lattices, Phys. Rev. Lett. {\bf 103}, 013901 (2009).\\
18. G. Roati, C. D'Errico, L. Fallani, M. Fattori, C. Fort, M. Zaccanti, G. Modugno, M. Modugno, and M. Inguscio, Anderson localization of a non-interacting Bose-Einstein condensate, Nature {\bf 453}, 895 (2008).\\
19. P. Wang, Y. Zheng, X. Chen, C. Huang, Y.V. Kartashov, L. Torner, V.V. Konotop, and F. Ye, 
Localization and delocalization of light in photonic Moir\'e lattices, 
Nature {\bf 577}, 42 (2020).\\
20. S. Longhi, Topological Phase Transition in non-Hermitian Quasicrystals, Phys. Rev. Lett. {\bf 122}, 237601 (2019).\\
21. S. Longhi, Metal-insulator phase transition in a non-Hermitian Aubry-Andr\'e-Harper model, Phys. Rev. B {\bf 100}, 125157 (2019).\\
22.  Q.-B. Zeng and Y. Xu, Winding numbers and generalized mobility edges in non-Hermitian systems, Phys. Rev. Research {\bf 2}, 033052
(2020).\\
23. X. Cai, Boundary-dependent self-dualities, winding numbers, and asymmetrical localization in non-Hermitian aperiodic one-dimensional models, Phys.
Rev. B {\bf 103}, 014201 (2021).\\
24. S. Weidemann, M. Kremer, S. Longhi and A. Szameit,
Topological triple phase transition in non-Hermitian Floquet quasicrystals, Nature {\bf 601}, 354 (2022).\\
25. Q. Lin, T. Li, L. Xiao, K. Wang, W. Yi, and P. Xue, 
Topological Phase Transitions And Mobility Edges in Non-Hermitian Quasicrystals, Phys. Rev. Lett. {\bf 129}, 113601 (2022).\\
26. D.E. Logan and P.G. Wolynes, Dephasing and Anderson localization in topologically disordered systems, Phys. Rev. B {\bf 36}, 4135 (1987).\\
27. D.A. Evensky, R.T. Scalettar, and P.G. Wolynes, Localization and Dephasing Effects in a Time-Dependent Anderson Hamiltonian,  J. Chem. Phys. {\bf 94}, 1149 (1990).\\
28. J. C. Flores, Diffusion in disordered systems under iterative measurement. Phys. Rev. B {\bf 60}, 30 (1999).\\
29. S. A. Gurvitz, Delocalization in the Anderson Model due to a Local Measurement, Phys. Rev. Lett. {\bf 85}, 812 (2000).\\
30. L.Levi, Y. Krivolapov, S. Fishman, and M. Segev, Hyper-transport of light and stochastic acceleration by evolving disorder, 
Nature Phys. {\bf 8}, 912 (2012).\\
31. Y. Krivolapov, L. Levi, S. Fishman, M. Segev, and M. Wilkinson, 
Super-diffusion in optical realizations of Anderson localization, New J. Phys. {\bf 14} 043047 (2012).\\
32. I. Yusipov, T. Laptyeva, S. Denisov, and M. Ivanchenko, Localization in Open Quantum Systems,
Phys. Rev. Lett. {\bf 118}, 070402  (2017).\\
33. S. Gopalakrishnan, K.R. Islam, and M. Knap, Noise-Induced Subdiffusion in Strongly Localized Quantum Systems, Phys. Rev. Lett. {\bf 119}, 046601 (2017).\\
34. Y. Rath and F. Mintert, Prominent interference peaks in the dephasing Anderson model,
Phys. Rev. Research {\bf 2}, 023161 (2020).\\
35. S. Longhi, Anderson Localization in Dissipative Lattices, Ann. Phys. (Berlin) {\bf 535}, 2200658 (2023).\\
36. K. Wang, X. Qiu, L. Xiao, X. Zhan, Z. Bian, W. Yi, and P. Xue, Simulating Dynamic Quantum Phase Transitions in Photonic Quantum Walks,
Phys. Rev. Lett. {\bf 122}, 020501 (2019).\\
37.  A. Regensburger, C. Bersch, B. Hinrichs, G. Onishchukov, A. Schreiber, C. Silberhorn, and U. Peschel, Photon propagation in a discrete fiber network: an interplay of coherence and losses, Phys. Rev. Lett. {\bf 107}, 233902 (2011).\\
38. M. Wimmer, A. Regensburger, M.-A. Miri, C. Bersch, D.N. Christodoulides, and U. Peschel, Observation of optical solitons in PT-symmetric lattices, Nat. Commun. {\bf 6}, 7782 (2015).\\
39. S. Wang, C. Qin, W. Liu, B. Wang, F. Zhou, H. Ye, L. Zhao, J. Dong, X. Zhang, S. Longhi, and P. Lu, High-order dynamic localization and tunable temporal cloaking in ac-electric-field driven synthetic lattices, Nature Commun. {\bf 13}, 7653 (2022).\\
40.S. Longhi, Inhibition of non-Hermitian topological phase transitions in sliding photonic quasicrystals, Opt. Lett. {\bf 48}, 6251 (2023).

\end{document}